\documentclass[preprint,showpacs,preprintnumbers,amsmath,amssymb,showkeys,pra]{revtex4}

\usepackage{epsf}
\usepackage{dcolumn}
\usepackage{bm}
\everymath{\displaystyle}
\begin{document}

\title{General dynamics of tensor polarization of particles and nuclei in external fields}

\author{Alexander J. Silenko}

\affiliation{Research Institute for
Nuclear Problems, Belarusian State University, Minsk 220030, Belarus,\\
Bogoliubov Laboratory of Theoretical Physics, Joint Institute for Nuclear Research,
Dubna 141980, Russia}  

\date{\today}

\begin {abstract}
The tensor polarization of particles and nuclei becomes constant in the coordinate system rotating with the same angular velocity as the spin and rotates in the lab frame with the above angular velocity. The general equation defining the time dependence of the tensor polarization is derived. An explicit form of dynamics of this polarization is found in the case when the initial polarization is axially symmetric.
\end{abstract}


\maketitle

\section{Introduction}\label{Introduction}

Polarized beams are often considered as a research tool to
study the fundamental interactions and to search for new physics. Unlike spin-1/2 fermions, particles and nuclei with spin $s\ge1$ possess tensor polarization.
This polarization is an important property of particles and nuclei and can be measured with a good accuracy. Investigations of dynamics of tensor polarization of light nuclei (e.g., the deuteron with spin 1) made it possible to discover \cite{Barexpr,Azhgirei}, to predict \cite{Bar3,BarBart}, and to describe \cite{PRC,BarHE,TechPhysLett} some new effects. However, these effects are conditioned either by quadratic in spin interactions of deuteron \cite{Barexpr,Azhgirei,Bar3,BarBart,PRC,BarHE} or by quantum beats in positronium \cite{TechPhysLett}. To get a basic understanding of polarization effects, it is necessary to consider \emph{linear} in spin interactions of \emph{tensor-polarized} particles and nuclei. While this problem is rather important, it is not much discussed. For all we know, the dynamics of tensor polarization in external fields has been considered in Ref. \cite{EvTensorPolnLee}. In this work, some examples of evolution of the tensor polarization have been investigated and some properties of the $5\times5$ spin transfer matrix describing this polarization have been considered. Evidently, the use of the $5\times5$ matrix is much more difficult than that of the corresponding $2\times2$ matrix for the vector polarization.

In the present work, we develop a simple general approach to a description of dynamics of the tensor polarization of particles and nuclei when this dynamics is caused by linear in spin interactions with external fields. The approach proposed allows us to couple dynamics of the tensor and vector polarizations and to derive general formulas defining an evolution of particles/nuclei polarization in external fields.


\section{General properties of spin dynamics}\label{dynamics}

As is known, the spin
rotation is exhaustively described with the polarization
vector $\bm P$ defined by
\begin{eqnarray}
P_i =\frac{<S_i>}{s}, ~~~ i=x,y,z. \label{eq1P}\end{eqnarray}
Here $S_i$ are corresponding spin matrices and $s$ is the spin
quantum number. Averages of spin operators are expressed by their convolutions with the wave function
$\Psi(t)$. Particles with spin $s\geq1$ also possess a tensor
polarization. Main characteristics of such a polarization are
specified by the polarization tensor $P_{ij}$, which is given by
\begin{eqnarray}
P_{ij} = \frac{3 <S_iS_j + S_jS_i>-2s(s+1)\delta_{ij}}{2s(2s -
1)}, ~~~ i,j=x,y,z. \label{eqPRC}\end{eqnarray} The polarization
tensor satisfies the conditions $P_{ij}=P_{ji}$ and
$P_{xx}+P_{yy}+P_{zz}=0$ and therefore has five independent
components. In the general case, the polarization vector and the
polarization tensor are time-dependent. Additional tensors
composed of products of three or more spin matrices are needed
only for the exhaustive description of polarization of
particles/nuclei with spin $s\ge3/2$.

The eight parameters defined by the polarization vector and the polarization tensor are independent.  In
particular, $<S_iS_j>\neq <S_i><S_j>$. 

The quantum-mechanical Hamiltonian describing the interaction of the spin with external fields contains 
terms linear and bilinear on the spin:
\begin{equation} {\cal H}=\bm\Omega\cdot\bm S+Q_{jk}S_jS_k.
\label{VveEtot} \end{equation} Here $\bm S$ is the matrix operator
describing the rest-frame spin. Since the spin matrices satisfy
the commutation relation
\begin{equation} [S_i,S_j]=ie_{ijk}S_k ~~~ (i,j,k=x,y,z),
\label{commuta} \end{equation} the first term in Eq.
(\ref{VveEtot}) defines the spin rotation with the angular
velocity $\bm\Omega$. The absolute value and the direction of
$\bm\Omega$ may arbitrarily depend on time.

Even when the second term in Eq. (\ref{VveEtot}) is omitted,
dynamics of the polarization tensor is not trivial. We cannot
characterize the particle spin wave function only by a single
frequency. For a spin-1 particle, there are two frequencies,
$+\Omega$ and $-\Omega$, defining three equidistant levels. For a
particle with the integer spin $s=N$, there are $2N$ frequencies,
$\pm\Omega$, $\pm2\Omega$, ..., $\pm N\Omega$, defining $2N+1$
equidistant levels. The single frequency of the spin rotation,
$\Omega$, originates from the specific commutation relations
(\ref{commuta}) for the spin operators $S_i$. For a spin-1
particle, this property of the spin operators manifests in the
fact that $(S_i)_{13}=(S_i)_{31}=0~(i=x,y,z)$. Therefore, the spin
operators mix only two neighboring levels. But it is not the case
for the tensor polarization operators. Three of them mix the
levels 1 and 3 because the components $(S_{x}^2)_
{13},~(S_{x}^2)_{31},~(S_{y}^2)_{13},~(S_{y}^2)_{31},~(S_{x}S_{y}+S_{y}S_{x})_{13}$,
and $(S_{x}S_{y}+S_{y}S_{x})_{31}$ are nonzero. As a result, the
frequency $2\Omega$ also appears. For particles with greater spins
($s\ge1$), the situation is still more complicated.

We consider electromagnetic interactions, namely,
coupling of spins to electric and magnetic fields.
The second term in Eq. (\ref{VveEtot}) characterizes the tensor
electric and magnetic polarizabilities and the quadrupole
interaction. It influences dynamics of both the polarization
vector and the polarization tensor. As a rule, this term is much
less than the first one. Its order of magnitude can be easily
determined. We can calculate the commutator $[{\cal H},S_i]$
taking into account that $s\sim1$. The term under consideration
brings a correction of the order of $|Q_{jk}S_jS_k|$ to the
angular frequency of spin precession $\Omega$. Let us evaluate
this correction whose effect on spin dynamics is rather complicate
and does not reduce to a change of $\bm\Omega$ (see Refs.
\cite{Bar3,BarBart,PRC}).
The angular frequency of spin precession of the deuteron in the magnetic field $B=3$ T is equal to $1.2\times10^8$ s$^{-1}$. If the tensor magnetic polarizability of the deuteron is close to the value $\beta_T=1.95\times10^{-40}$ cm$^3$ predicted by the theory \cite{CGS,JL}, the related \emph{effective} change of the angular frequency of spin precession is about $2\times10^{-4}$ s$^{-1}$. 
The predicted values of the tensor electric polarizability of the deuteron \cite{CGS,JL,FP} are slightly less and cannot lead to a greater effect. On the other hand, spin-tensor effects can be caused by intrinsic electric quadrupole moments of nuclei. For example, a spatial derivative of an electric field strength in Penning traps is usually of the order of $10^6$ V/m$^2$. The quadrupole moment of the deuteron is equal to $Q_d=0.286$  fm$^2$. A simple estimate shows that the related \emph{effective} change of the angular frequency of spin precession of the deuteron is of the order of $10^{-8}$ s$^{-1}$. For polarized deuteron beams in accelerators and storage rings, the intrinsic electric quadrupole moment 
also causes spin-tensor interactions. However, its effect on the spin evolution
is a few orders of magnitude lower than that of the tensor magnetic and electric polarizabilities.
When $B=3$ T, the ring radius $R\sim10$ m, and the field index $n=-(R/B)(\partial B/\partial r)\sim0.1$, the \emph{effective} change of the angular frequency of spin precession caused by the electric quadrupole moment 
is of the order of $10^{-9}$ s$^{-1}$ even for relativistic deuterons. 

Thus, one can usually neglect an influence of the spin-tensor interactions on spin dynamics. Such an influence can be significant only in special cases like hyperfine interactions in atoms or in experiments specially designed to measure the tensor magnetic and electric polarizabilities. Hereafter, we will disregard the spin-tensor interactions and will take into account only the first term in the interaction Hamiltonian (\ref{VveEtot}).

The equation of spin motion is given by
\begin{equation} \frac{d\bm S}{dt}=\frac i\hbar[{\cal H},\bm S]=\bm\Omega\times\bm S.
\label{VveEfft} \end{equation} When the particle/nucleus moves in electromagnetic fields, $\bm\Omega$ is defined by the Thomas-Bargmann-Michel-Telegdi (T-BMT) equation \cite{T-BMT} extended on the electric dipole moment (EDM):
\begin{equation} \begin{array} {c}
\bm \Omega=-\frac{e}{mc}\left[\left(G+\frac{1}{\gamma}\right){\bm B}-\frac{\gamma G}{\gamma+1}({\bm\beta}\cdot{\bm B}){\bm\beta}-\left(G+\frac{1}{\gamma+1}\right)\bm\beta\times{\bm E}\right.\\
+\left.\frac{\eta}{2}\left({\bm E}-\frac{\gamma}{\gamma+1}(\bm\beta\cdot{\bm E})\bm\beta+\bm\beta\times {\bm B}\right)\right].
\end{array} \label{Nelsonh} \end{equation}
Here $G=(g-2)/2,~g=2mc\mu/(e\hbar s),~\eta=2mcd/(e\hbar s),~\bm\beta={\bf v}/c$, $\gamma$ is the Lorentz factor, and $\mu$ and $d$ are the magnetic and electric dipole moments.
An extension of the T-BMT equation due to the EDM has already been discussed in the original paper of Bargmann, Michel and Telegdi \cite{T-BMT}. Then, the equation of spin motion of the particle with the anomalous magnetic moment $\mu'$ and the EDM has been obtained in Refs. \cite{Nelson,Khriplovich} by the dual
transformation
$\mu'\rightarrow d,~{\bm B}\rightarrow{\bm E},~{\bm E}\rightarrow-{\bm B}$.
The rigorous derivation of this equation has been presented in Ref. \cite{FukuyamaSilenko}. The resulting equation of spin motion coincides with that presented in Refs. \cite{Nelson,Khriplovich}. However, the derivation fulfilled in Ref. \cite{FukuyamaSilenko} has not used the supplementary assumption of dual symmetry.

The Pauli matrices together with the unit matrix describe the spin of a Dirac particle and generate an
irreducible representation of the $SU(2)$ group.
Algebraically, the $SU(2)$ group is a double cover of the
three-dimensional rotation group $SO(3)$. As a result, the spin dynamics defined by the Dirac equation fully corresponds to the classical picture of rotation of an intrinsic angular momentum (spin) in external fields. The angular velocity of spin rotation does not depend on the spin quantum number. Therefore, both the classical description and the formalism
based on the Pauli matrices are applicable to particles/nuclei with an
arbitrary spin if the effect of spin rotation is analyzed.
Certainly, this formalism becomes insufficient if one considers spin-tensor effects mentioned in Sec. \ref{Introduction}.

In Eq. (\ref{Nelsonh}), we can estimate terms dependent on the EDM. Any experimental constraints on the deuteron EDM are not established yet. For the proton, the current indirect bound of
$|d_p| < 7.9\times10^{-25}$ e$\cdot$cm has been obtained using Hg atoms \cite{Griffith}. If the deuteron EDM was equal to this value, its contribution to the angular frequency of spin precession of 1 GeV/c deuterons in the magnetic field $B=3$ T were equal to $5\times10^{-3}$ s$^{-1}$.
This means that one can often neglect an influence of the spin-tensor terms in the Hamiltonian (\ref{VveEtot}) on the spin dynamics 
but can keep the EDM-dependent terms in Eq. (\ref{Nelsonh}). Certainly, a possible future ascertainment of very strong restrictions on the deuteron EDM can change this situation.

\section{Huang-Lee-Ratner approach}\label{HuangLee}

Let us consider the approach used by Huang, Lee, and Ratner \cite{EvTensorPolnLee} to describe an evolution of the tensor polarization.
Instead of the polarization tensor (\ref{eqPRC}), the following spin-tensor operators can be used \cite{EvTensorPolnLee}:
\begin{equation} \begin{array} {c}
T_0=\frac{1}{\sqrt2}\left(3S_z^2-2\right), ~~~T_{\pm1}=\pm\frac{\sqrt3}{2}\left(S_\pm S_3+S_3S_\pm\right), ~~~ T_{\pm2}=\frac{\sqrt3}{2}S_\pm^2.
\end{array} \label{TensrPolnLee} \end{equation}
To obtain observable quantities, these operators should be averaged.

The approach proposed in Ref. \cite{EvTensorPolnLee} was based on the T-BMT equation and disregarded the spin-tensor interactions in the Hamiltonian (\ref{VveEtot}). This approach used the the Frenet-Serret curvilinear coordinates. Expressing the spin vector in terms of its
components,
\begin{equation} \begin{array} {c}
\bm S=S_1 \bm e_1+S_2 \bm e_2+S_3 \bm e_3,
\end{array} \label{Lee} \end{equation}
and defining $S_{\pm} = S_1\pm iS_2$, one obtains the following equation of spin motion:
\begin{equation} \begin{array} {c}
\frac{dS_{\pm}}{d\phi}=\pm iG\gamma S_{\pm}\pm iF_\pm S_3, ~~~ \frac{dS_{3}}{d\phi}=\frac i2\left(F_- S_+- F_+ S_-\right),
\end{array} \label{Leet} \end{equation}
where $F_\pm$ characterize the spin
depolarization kick and $\phi$ is the azimuthal orbit rotation angle. The spin tune, $G\gamma$, is the number of spin revolutions per orbit turn.

Equations (\ref{TensrPolnLee})--(\ref{Leet}) allow one to obtain the set of equations defining the evolution of the tensor polarization \cite{EvTensorPolnLee}:
\begin{equation} \begin{array} {c}
\frac{dT_0}{d\phi}=\frac{3}{\sqrt2}\left(\frac{dS_{3}}{d\phi}S_{3}+S_{3}\frac{dS_{3}}{d\phi}\right)=-i\frac{\sqrt6}{2}\left(F_- T_{+1}+ F_+ T_{-1}\right), \\
\frac{dT_{\pm1}}{d\phi}=-\frac{\sqrt3}{2}\left(\frac{dS_{\pm}}{d\phi}S_{3}+S_{3}\frac{dS_{\pm}}{d\phi}+\frac{dS_{3}}{d\phi}S_{\pm}+S_{\pm}\frac{dS_{3}}{d\phi}
\right)=\pm iG\gamma T_{\pm1}-i\frac{\sqrt6}{2}F_\pm T_{0}- iF_\mp T_{\pm2}, \\
\frac{dT_{\pm2}}{d\phi}=\frac{\sqrt3}{2}\left(\frac{dS_{\pm}}{d\phi}S_{\pm}+S_{\pm}\frac{dS_{\pm}}{d\phi}\right)=\pm 2iG\gamma T_{\pm2}-iF_\pm T_{\pm1}.
\end{array} \label{TPLee} \end{equation}
These equations can be presented in the matrix form:
\begin{equation} \begin{array} {c}
\frac{dT}{d\phi}=AT,\\ T=\left(\begin{array}{c}T_{+2}\\T_{+1}\\T_0\\T_{-1}\\T_{-2}\end{array}\right), ~~~
A=\left(\begin{array}{ccccc}2iG\gamma&-iF_+&0&0&0\\-iF_-&iG\gamma&-i\frac{\sqrt6}{2}F_+&0&0\\
0&-i\frac{\sqrt6}{2}F_-&0&-i\frac{\sqrt6}{2}F_+&0\\0&0&-i\frac{\sqrt6}{2}F_-&-iG\gamma&-iF_+\\
0&0&0&-iF_-&-2iG\gamma
\end{array}\right).
\end{array} \label{Leema} \end{equation}

In Ref. \cite{EvTensorPolnLee}, some applications of this approach have been considered.

While the approach of Huang, Lee, and Ratner is absolutely correct, we can make some evident remarks. First of all, the use of $5\times5$ matrices instead of standard $2\times2$ accelerator matrices needs much more cumbersome derivations. Moreover, even investigations of spin-tensor effects performed in Ref. \cite{PRC} used three-component wave functions and $3\times3$ matrices. Another remark is an absence of a direct connection between evolutions of the polarization vector and the polarization tensor, while spin-tensor interactions are not taken into consideration in Ref. \cite{EvTensorPolnLee}.

In the present work, we propose a different approach which couples dynamics of
the tensor and vector polarizations and simplifies a description of the evolution of the
tensor polarization of partcles/nuclei in external fields.

\section{Evolution of the polarization tensor}\label{vectorpol}

It is not evident whether the evolution of the polarization tensor also reduces to the rotation. To clear the problem, we use the following approach. Let us consider the Cartesian coordinate system rotating about the $z$ axis with the same angular velocity $\bm\Omega(t)$ as the spin. In the general case, this angular velocity depends on time. We may superpose the rotating and nonrotating coordinate systems at the initial moment of time $t=0$ ($\bm e'_i(0)=\bm e_i$). The rotating coordinate system is denoted by primes. 


The spin components in the rotating coordinate system remain unchanged:
\begin{equation} \frac{dS'_i}{dt}=0~~~(i=x,y,z).
\label{reunc} \end{equation}

As a result, all tensor polarization operators and all components of the polarization tensor are also unchanged in this coordinate system ($S'_iS'_j+S'_jS'_i=const,~P'_{ij}=const$). This important property shows that the tensor polarization of particles/nuclei with spin $s\ge1$ rotates in external fields similarly to the vector polarization. This is valid not only for electromagnetic interaction but also for other (weak, gravitational) interactions. Other products of the spin operators $S'_iS'_j\dots S'_k$ are also conserved in the rotating coordinate system.

To determine the particle polarization in the nonrotating coordinate system (lab frame), we need to connect the directions of basic vectors of the primed and unprimed coordinate systems. This problem can be easily solved. We choose the initial moment of time $t=0$. Evolution of any basic vector of the primed coordinate system coincides with that of the spin when it is initially directed along the considered basic vector. Therefore, the final directions of the three primed basic vectors are defined by the three final directions of the spin when its corresponding initial ones are parallel to the three Cartesian axes of the unprimed coordinate system.
As a result, dynamics of the polarization tensor reduces to that of the polarization vector. In particular, the evolution of the polarization tensor in accelerators and storage rings can be unambiguously defined with usual $2\times2$ spin transfer matrices. Thus, one need not apply more cumbersome $5\times5$ matrices.

Time dependence of the polarization tensor defined in the lab frame can be easily expressed in terms of the basic vectors $\bm e'_i$. 
Since $\bm e'_i(0)=\bm e_i$ and
\begin{eqnarray}
S'_i(t)=\sum_{k}{(\bm e'_i(t)\cdot\bm e_k)S_k},
\label{eqppnn}\end{eqnarray} the polarization tensor is given by
\begin{eqnarray}
P_{ij}(t) = \frac{3}{2s(2s - 1)}\sum_{k,l}{\left[(\bm e'_i(t)\cdot\bm e_k)(\bm e'_j(t)\cdot\bm e_l) <S_kS_l + S_lS_k>\right]}- \frac{s+1}{2s - 1}\delta_{ij}. \label{peqptfn}\end{eqnarray}

This simple equation defines general dynamics of the tensor polarization of particles and nuclei in
external fields.

\section{Evolution of axially symmetric polarization}

A calculation of the polarization tensor can usually be still more simplified. In particular, this is possible when the initial polarization is axially symmetric. It is convenient to direct the $z'$ axis along the symmetry axis. 
For the considered initial polarization, $P'_{xx}=P'_{yy}$ and $P'_{ij}=0$ when $i\neq j$. It has been ascertained in the precedent section that the primed polarization tensor remains unchanged.

Let us suppose that the direction of $\bm e'_z(t)$ in the lab frame is defined by the time-dependent spherical angles $\theta$ and $\phi$. In the considered case, the lab frame polarization tensor does not depend on directions of the $x'$ and $y'$ axes in the plane orthogonal to $z'$ and we can choose the primed basic vectors as follows:
\begin{equation}\begin{array}{c}
\bm e'_x=\bm e_x \cos{\theta}\cos{\phi}+\bm e_y \cos{\theta}\sin{\phi}-\bm e_z \sin{\theta},~~~ \bm e'_y=-\bm e_x \sin{\phi}+\bm e_y \cos{\phi},\\ \bm e'_z=\bm e_x \sin{\theta}\cos{\phi}+\bm e_y \sin{\theta}\sin{\phi}+\bm e_z \cos{\theta}.
\end{array}\label{pribave} \end{equation}

The connection between the spin operators in the unprimed and primed coordinate systems is defined by Eqs. (\ref{eqppnn}) and (\ref{pribave}) and is given by
\begin{equation}\begin{array}{c}
S_x=S'_x \cos{\theta}\cos{\phi}-S'_y \sin{\phi}+S'_z \sin{\theta}\cos{\phi},~~~ S_y=S'_x \cos{\theta}\sin{\phi}+S'_y \cos{\phi}+S'_z \sin{\theta}\sin{\phi},\\ S_z=-S'_x \sin{\theta}+S'_z \cos{\theta}.
\end{array}\label{pribapn} \end{equation} Due to the axial symmetry, averaging vanishes $<S'_x>$ and $<S'_y>$. The polarization vector $\bm P=<\bm S>/s$ is given by
\begin{equation}\begin{array}{c}
P_x(t)=P_z(0) \sin{\theta}\cos{\phi},~~~ P_y(t)=P_z(0)\sin{\theta}\sin{\phi},~~~ P_z(t)=P_z(0)\cos{\theta}.
\end{array}\label{pribatn} \end{equation}

The use of Eqs. (\ref{peqptfn}) and (\ref{pribapn}) leads to the following equation defining the polarization tensor:
\begin{equation}\begin{array}{c}
P_{xx}(t)=A\left(\sin^2{\theta}\cos^2{\phi}-\frac13\right), ~~~P_{yy}(t)=A\left(\sin^2{\theta}\sin^2{\phi}-\frac13\right), \\ P_{zz}(t)=A\left(\cos^2{\theta}-\frac13\right), ~~~
P_{xy}(t)=A\sin^2{\theta}\sin{\phi}\cos{\phi}, \\ P_{xz}(t)=A\sin{\theta}\cos{\theta}\cos{\phi}, ~~~P_{yz}(t)=A\sin{\theta}\cos{\theta}\sin{\phi}, ~~~ A=\frac32P_{zz}(0).
\end{array}\label{pribapt} \end{equation}

This general equation exhaustively describes dynamics of the tensor polarization of particles and nuclei in external fields. The time dependence of the angles $\theta$ and $\phi$ can be determined with the use of the spin motion equations (\ref{VveEfft}) and (\ref{Nelsonh}). For particles/nuclei in accelerators and storage rings, $2\times2$ spin transfer matrices are commonly applied \cite{Lee,MZ}. To take into account spin-tensor interactions transforming the tensor polarization into the vector one and the other way
round, $3\times3$ spin matrices should be used (see Ref. \cite{PRC}).

\section{Summary}

In the present work, we have considered the evolution of the tensor polarization of particles and nuclei when this evolution is caused by linear in spin interactions with external fields. The tensor polarization becomes constant in the coordinate system rotating with the same angular velocity as the spin. This means that the tensor polarization rotates in the lab frame with the above angular velocity. We have derived the general equation (\ref{peqptfn}) defining the time dependence of this polarization. Calculations becomes appreciably simpler when the initial polarization is axially symmetric. In this case, the explicit form of dynamics of the tensor polarization is given by Eq. (\ref{pribapt}).

The approach used has allowed us to couple dynamics of the tensor and vector polarizations. This result considerably simplifies a determination of the tensor polarization because the vector one is
exhaustively described by $2\times2$ spin transfer matrices. As a result, one need not apply $5\times5$ matrices to investigate the evolution of five independent components of the polarization tensor. The obtained general equations can be widely used (e.g., for polarized beams in accelerators and storage rings).

\section*{Acknowledgements}

The work was supported in part by the Belarusian Republican Foundation for
Fundamental Research (Grant No. $\Phi$14D-007) and by the Heisenberg-Landau program of the German Ministry for Science and Technology (Bundesministerium f\"{u}r Bildung und Forschung).


\end{document}